\documentclass[11pt,3p,twocolumn]{elsarticle}

\journal{arXiv}

\usepackage[utf8]{inputenc}		%
\usepackage[T1]{fontenc}		%
\usepackage{graphicx}			%
\usepackage{amssymb} 			% Symbols
\usepackage{amsmath} 			% Symbols
\usepackage{wasysym}
\usepackage{lineno}
\begin{document}

\begin{frontmatter}

\title{Ensuring Uninterrupted Power Supply to Lunar Installations Through an Organic Rankine Cycle}

\author{F. Francisco}
\ead{frederico.francisco@fc.up.pt}

\author{O. Bertolami}
\ead{orfeu.bertolami@fc.up.pt}

\address{Departamento de Física e Astronomia and Centro de Física do Porto, Faculdade de Ciências, Universidade do Porto, Rua do Campo Alegre 687, 4169-007 Porto, Portugal}

\begin{abstract}
	We propose using the temperature gradients between the Moon's surface and the soil at a certain depth to power an Organic Rankine Cycle that could supply a permanent installation, particularly at night, when solar power is not available. Our theoretical and engineering considerations show that, with existing working fluids and quite feasible technical requirements, it is possible to continuously yield $25\,{\rm kW}$ to sustain a 3 member crew.
\end{abstract}


\end{frontmatter}

%-----------------------------------------------------------------------------%
%-- Body ---------------------------------------------------------------------%
%-----------------------------------------------------------------------------%

\section{Introduction}

The celebration of the fifty years of the human landing on the moon should be regarded as a moment for reflection and understanding of the urgency to gather efforts and resources to seriously consider the steps towards its colonization and systematic scientific exploration. The Moon is the most immediate and consensual cosmic challenge and its colonization will boost the development of critical technological hurdles for further and continuous space exploration.

Indeed, the Moon can unravel unique opportunities for science, engineering and resource exploitation. As an example of scientific and engineering goals, we could mention the advantageous opportunities for further lunar laser ranging \cite{Dickey:1994,Williams:2004}, astronomy, due to the inexistence of an atmosphere, and radio-astronomy \cite{Wolt:2012,Silk:2018}. It is also well know that Moon's soil is particularly rich in $\rm He^3$, the fuel of the future fusion nuclear reactors \cite{Heiken:1991}. Furthermore, the impact of comets and asteroids on the Moon are important sources of metals, ice and compounds that can supply humankind for many centuries (see Ref.\,\cite{Heiken:1991} for an extensive review on the Moon).

In fact, the colonization of space presents a number of technical challenges, most of which are yet to be overcome. One of which is how to ensure a sustained power supply given the specific conditions of the Moon.

The power requirements of an initial lunar base camp with 3 crew members have been estimated at $25\,{\rm kW}$. However an advanced base with an industrial or mining operation could need over $1\,{\rm MW}$ \cite{Landis:1990}.

A key feature of the lunar environment when considering long term settlement is the length of the day-night cycles. The moon has a rotation period of approximately 27 days, which is tidally coupled with its orbital period around Earth. This means each day and night on the Moon lasts for approximately 14 Earth days. This poses a problem regarding a continuous power supply.

The thermal amplitudes on the Moon are extreme, oscillating between day and night mean temperatures of approximately $380\,{\rm K}$ and $120\,{\rm K}$ ($107^{\circ}{\rm C}$ and $-153^{\circ}{\rm C}$), respectively \cite{Heiken:1991}.

\begin{figure}
	\centering
	\includegraphics[width=\columnwidth]{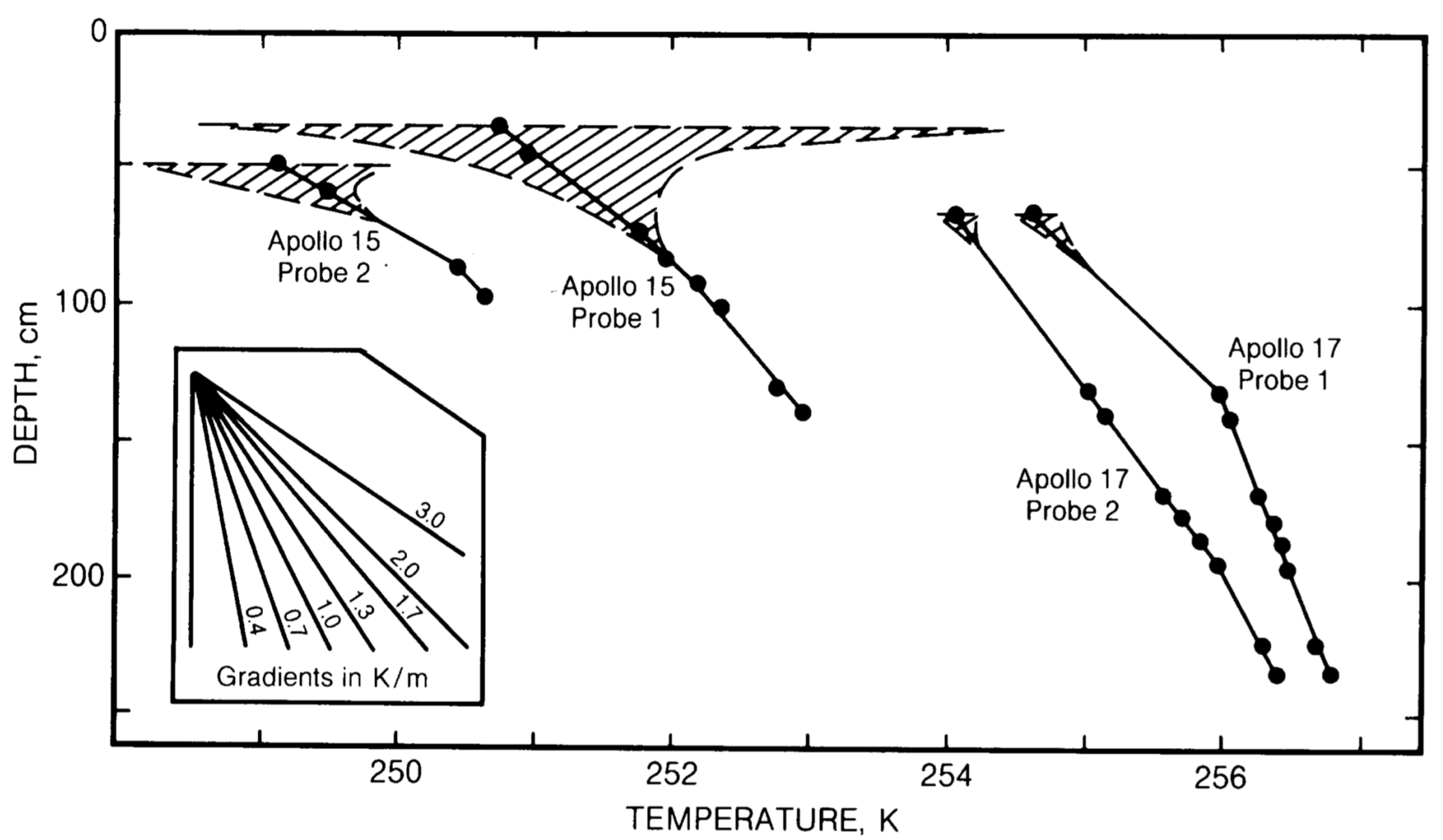}
	\caption{Lunar regolith temperature profile from measurements conducted by Apollo austronauts. Adapted from Ref.\,\cite{Heiken:1991}.}
	\label{fig:RegolithTemperatures}
\end{figure}

Owing to these sharp temperature amplitudes, the lunar soil has significant temperature gradients with depth. Measurements conducted during the Apollo lunar landings show that at depths below $80~{\mathrm cm}$ in the lunar regolith the day-night temperature variations are no longer present. Even below $50~{\mathrm cm}$ the temperature fluctuations are on the order of only $\pm 1~{\mathrm K}$. The mean soil temperatures in the first few meters where the measurements were taken are on the order of $250~{\mathrm K}$ and there is a temperature gradient of the order of $1$ to $3~{\rm K/m}$ (\textit{ct.} Figure\,\ref{fig:RegolithTemperatures}). The gradient arises from the heat flow from the Moon's crust \cite{Heiken:1991}.

The default option for supplying power to a lunar colony has been to use photovoltaic cells in solar panels, similarly to most power systems in space. The use of solar panels as the exclusive power source, requires some kind of energy storage to cover for periods in the shadow. However, current battery technology is far from being able to cover for 14 days without sunlight, unless massive battery farms were to be used.

There have been proposals to store thermal energy in a heat mass made of processed lunar regolith. In this concept, a solar power concentrator heats a mass of compacted lunar regolith to high temperatures. The stored heat is then used during the night to power a Stirling engine that produces electricity \cite{Climent:2014,Lu:2016}.

The presence of the aforementioned temperature gradients in the Moon's soil presents an opportunity to build a thermodynamic power system that can ensure the long-term continuity of power supply on the Moon. Our proposal is to use these temperature gradients to power a thermal engine to supply an installation with uninterrupted power. The main issue when considering classic thermal power systems, usually based on water or air as working fluid, is that they require very high temperatures only achievable either by burning some kind of fuel or through concentration of solar power.

Recently, however, there has been an increasing number of proposals on the application of what is usually called an Organic Rankine Cycle (ORC), which has been proposed as an alternative for power generation from low-temperature heat sources such as solar heat, waste heat or geothermal energy \cite{Yamamoto:2001}. This cycle is characterized by the use of an organic working fluid instead of water, allowing for heat conversion in low-temperature sources \cite{Yamamoto:2001,Saleh:2007}. The organic working fluids typically have much lower melting and boiling temperatures than water, allowing the engine to work at lower heat source temperatures. Many of the working fluids employed are the ones commonly used in refrigeration cycles. Current applications of ORC on Earth include cogeneration facilities \cite{Oyewunmi:2017}, Ocean Thermal Energy Conversion \cite{Yamada:2009} and low-grade geothermal heat sources \cite{MadhawaHettiarachchi:2007}.

In this paper, we propose that an ORC can be used effectively as the basis of a thermal power system using the temperature gradients in the Moon's soil, particularly, during the long lunar night. As we explore a set of possible working fluids, we examine the feasibility of an ORC to power a lunar colony during its long nights and estimate its performance.

%-----------------------------------------------------------------------------%
%-----------------------------------------------------------------------------%

\section{The Rankine Cycle}

A Rankine Cycle is a closed thermodynamic cycle where the conversion of heat into mechanical work is done through a liquid-gas phase change of the working fluid. It is one of the standard thermodynamic cycles presented in most textbooks on the subject \cite{Moran:2014}. The standard steam Rankine Cycle has a ubiquitous application in steam-based power systems, from steam locomotives, to coal-fired or nuclear power plants. It is worth reviewing the thermodynamics of this cycle.

\begin{figure}
	\centering
	\includegraphics[width=\columnwidth]{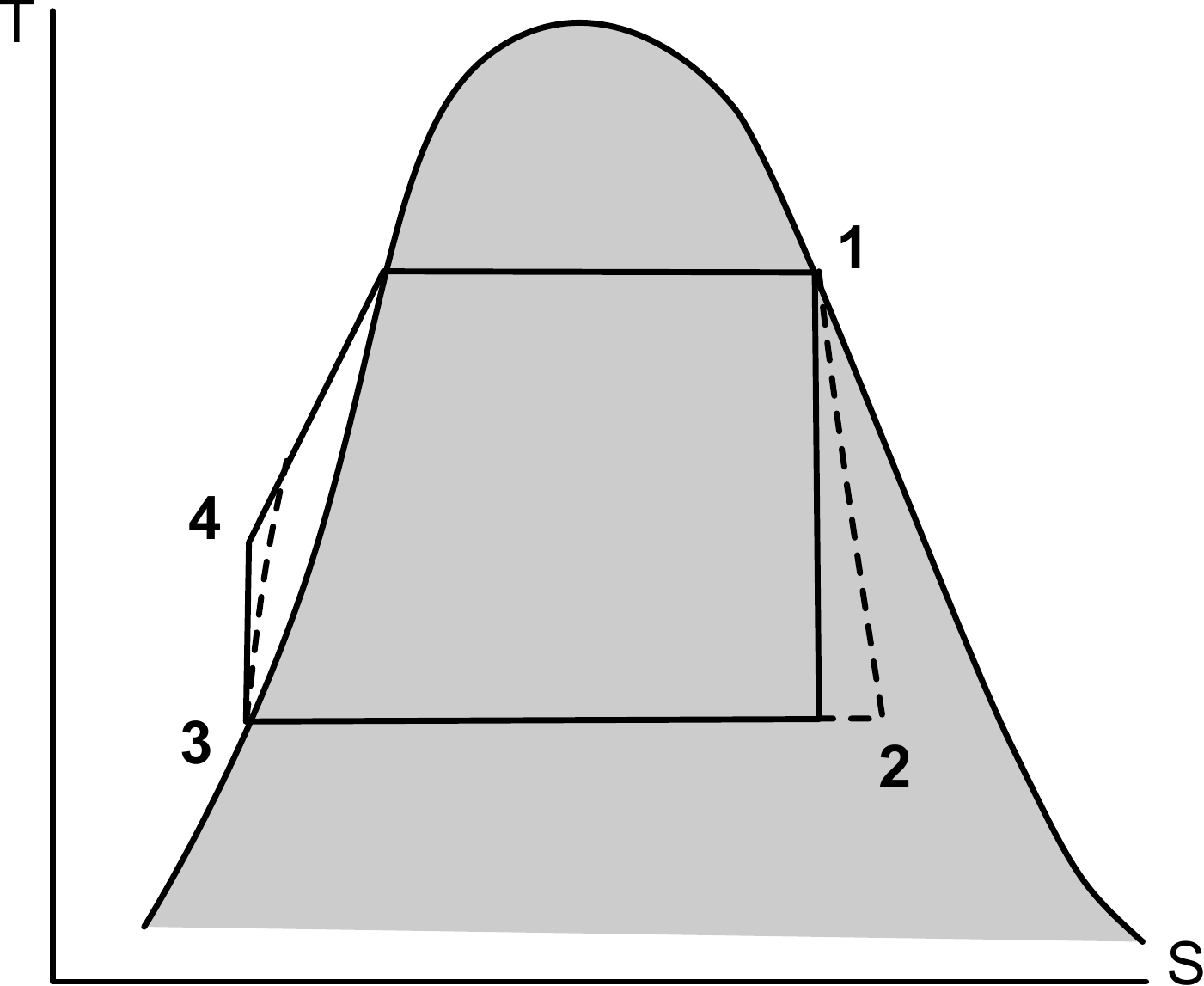}
	\caption{Temperature-entropy (T-S) diagram schematic of a Rankine cycle. Dashed lines represnd the effect of irreversibilities in the turbine and pump.}
	\label{fig:Rankine_cycle}
\end{figure}

In a Rankine cycle, the working fluid is first heated until it becomes saturated vapour. This corresponds to point (1) in the $T-S$ diagram in Figure\,\ref{fig:Rankine_cycle}. That vapour is then fed into a turbine that expands it and converts thermal energy into mechanical work. In an ideal turbine, the conversion is isentropic and thus
\begin{equation}
	{\dot{W}_{\rm t,is} \over \dot{m}} = h_1 - h_{2\text{,is}},
\end{equation}
where $\dot{W}_{\rm t,is}$ is the isentropic rate of work supplied by the turbine into the shaft, $\dot{m}$ is the mass flow and $h_1,h_2$ are the specific enthalpies of the fluid before and after passing through the turbine, respectively. In order to account for the effect of irreversibilities in the turbine, we must define the turbine efficiency
\begin{equation}
	\eta_{\rm t} = {\dot{W}_{\rm t} \over \dot{W}_{\rm t,is}} = {h_1 - h_2 \over h_1 - h_{2\text{,is}}},
\end{equation}
where $\dot{W}_{\rm t}$ is the actual work rate supplied by the turbine.

The fluid is then cooled in a condenser, transferring heat at a rate given by
\begin{equation}
	{\dot{Q}_{\rm out} \over \dot{m}} = h_2 - h_3
\end{equation}
to the cold source, until it becomes a saturated liquid. The enthalpy must then be raised by a pump so that it is fed into the hot source at the correct pressure to become a saturated gas when leaving it, thus closing the cycle. The isentropic work supplied by the pump is given by
\begin{equation}
	{\dot{W}_{\rm p} \over \dot{m}} = h_{4\text{,is}} - h_2,
\end{equation}
with the pump efficiency defined, similarly to the case of the turbine, as
\begin{equation}
	\eta_{\rm p} = {\dot{W}_{\rm p} \over \dot{W}_{\rm p,is}} = {h_4 - h_3 \over h_4 - h_{3\text{,is}}},
\end{equation}

Finally, the heat transferred to the fluid by the hot source is
\begin{equation}
	{\dot{Q}_{\rm in} \over \dot{m}} = h_1 - h_4.
\end{equation}

The two design parameters for an ideal Rankine cycle are the temperatures of the hot and cold sources, $T_{\rm H}$ and $T_{\rm C}$, which, ideally, correspond to the turbine inlet temperature, $T_1$, and condenser exit temperature, $T_3$, respectively. 

With these temperatures, we can define the evaporator and condenser pressures, $p_{\rm boil} = p_1 = p_4$ and $p_{\rm cond} = p_2 = p_3$, respectively, since heat transfer is made at constant pressure. Assuming it is a saturated vapour Rankine cycle, the specific enthalpies, $h_1$ and $h_3$, and specific entropies, $s_1$ and $s_3$, of the states (1) and (3) are also unambiguously defined for a given working fluid. Then, by assuming that the pump and turbine are ideal, we can find $s_2 = s_1$ and $s_4 = s_3$, thus completely defining the ideal cycle. 

After accounting for the turbine and the pump inefficiencies, the overall thermal efficiency of the cycle is given by
\begin{equation}
	\eta = {\dot{W}_{\rm t} \over \dot{Q}_{\rm in}} = 1 - {h_2 - h_3 \over h_1 - h_4}.
\end{equation}

Since all expressions are in terms of the mass flow rate, $\dot{m}$, this will be determined by the total power requirement set for the engine.

%-----------------------------------------------------------------------------%
%-----------------------------------------------------------------------------%

\section{Ideal Cycle Design}

Our proposal is to use the heat gradients present in the Moon's soil to feed an engine with thermal energy. This is, at least theoretically, viable given the significant temperature differences one can find with only a few meters in depth,

As in any thermal engine, the farther apart are the hot and cold source temperatures, the higher the thermal efficiency. However, in this case, there is always at least one temperature over which no control can be exercised, the surface temperature, while the other can only be controlled by changing the depth at which heat is transferred. Still, the temperature gradients after the first meter, where the day-night oscillations are no longer felt, are of only $1\sim2\,{\rm K/m}$.

Let us then, for the purpose of this analysis, assume that heat can be transferred to or from the regolith at a depth of a few meters at a temperature of approximately $250~{\rm K}$.

Since temperatures at the surface have such a wide range between night and day, it would act as cold source during the night and hot source during the day. In this case, either a reversible thermal engine or two separate machines would be needed if intended for permanent use. Alternatively, conventional photovoltaic power could be used during the day and use an ORC thermal engine to ensure continuity of supply during the lunar night.

If we design our ORC to work during the lunar night, then a surface temperature close to $120~{\rm K}$ sets the limit on the cold source, if conductive heat dissipation to the surface soil is used. Given the extremely low atmospheric densities, convective heat transfer is out of the question. The use of a radiator, however, could allow for temperatures even lower than the surface soil temperature if the radiator can be shielded from radiation coming from the soil itself. However, as we shall see, the lower temperature limit is mostly inconsequential, since for the candidate working fluids it would require a pressure well bellow what is considered feasible.

The criteria for the selection of the working fluid are set by the temperatures of the hot and cold sources. Specifically, the boiling point should be below the maximum temperature of the cycle and the melting point should be below the minimum temperature. Thus, we need to find working fluids that can be used in an ORC with temperatures in that range from $120~{\rm K}$ to $250~{\rm K}$.

\begin{table*}
	\centering
	\caption{List of possible working fluids with possible performance figures for the ideal ORC with $T_{\rm H} = 250\,{\rm K}$ and $T_{\rm C} = 120\,{\rm K}$, where $p_{\rm boil}$ and $p_{\rm cond}$ are the boiler and condenser pressures, respectively, $\eta$ is the cycle thermal efficiency and $\dot{W}_{\rm t}/\dot{m}$ is the cycle rate of work per unit mass flow rate.}
	\label{tbl:working_fluids}
	\begin{tabular}{l c | c c | c c}
		Fluid	& Formula			& $p_{\rm boil}$& $p_{\rm cond}$& $\eta$	& $\dot{W}_{\rm t}/\dot{m}$	\\
		~		& ~					& $\rm (kPa)$	& $\rm (kPa)$	&			& $\rm (kJ/kg)$ \\
		\hline
		\hline
		R13		& $\rm CClF_3$		& $1040$		& $0.133$		& $42.5\%$	& $98.4$ \\
		R23		& $\rm CHF_3$		& $1263$		& $0.0819$		& $42.9\%$	& $145$ \\
		R41		& $\rm CH_3F$		& $-$			& $-$			& $-$		& $-$ \\
		R116 	& $\rm C_2F_6$		& $-$			& $-$			& $-$		& $-$ \\
		R170 (Ethane)& $\rm C_2H_6$	& $1300$		& $0.352$		& $43.3\%$	& $303$ \\
	\end{tabular}\\
\end{table*}

We have set up possible candidates in Table\,\ref{tbl:working_fluids}, most of them are fluids often used in refrigeration cycles, due to their low fusion and boiling temperatures. Still, not all candidates can attain the low temperatures of the cold source, and even those who do are at extremely low pressures.

The ideal cycle results for the performance of the ORC with the working fluids compiled in Table\,\ref{tbl:working_fluids} allow for a figure of merit of the fundamental limits for the application of an ORC on the Moon in the way we are proposing. The cycle thermal efficiency would be in the vicinity of $40\%$, if a working fluid could be found that works closer to the temperature limits. This compares with a Carnot cycle efficiency of $52\%$. However, the net work per mass flow rate has a much wider range, affecting the amount of work fluid that is required to obtain a certain amount of power and the size of the engine itself and its components.

%-----------------------------------------------------------------------------%
%-----------------------------------------------------------------------------%

\section{Engineering Considerations}

%-----------------------------------------------------------------------------%

\subsection{Evaporation and Condensation Pressure}

As mentioned above, there are practical limitations on the boiler and condenser pressures. Most references state a boiler pressure limit of $2\,{\rm MPa}$ and a minimum condenser pressure of the order of $0.1\,{\rm MPa}$ \cite{Bao:2013,Stijepovic:2012,Aljundi:2011}.

We shall use these limits as design criteria for our proposed ORC. As seen in Table\,\ref{tbl:working_fluids}, the boiler pressure figures are all well below the $2\,{\rm MPa}$ limit, so no problems arise here. However, the condenser pressures required for the cold source temperature are too low.

\begin{figure*}
	\centering
	\includegraphics[width=\columnwidth]{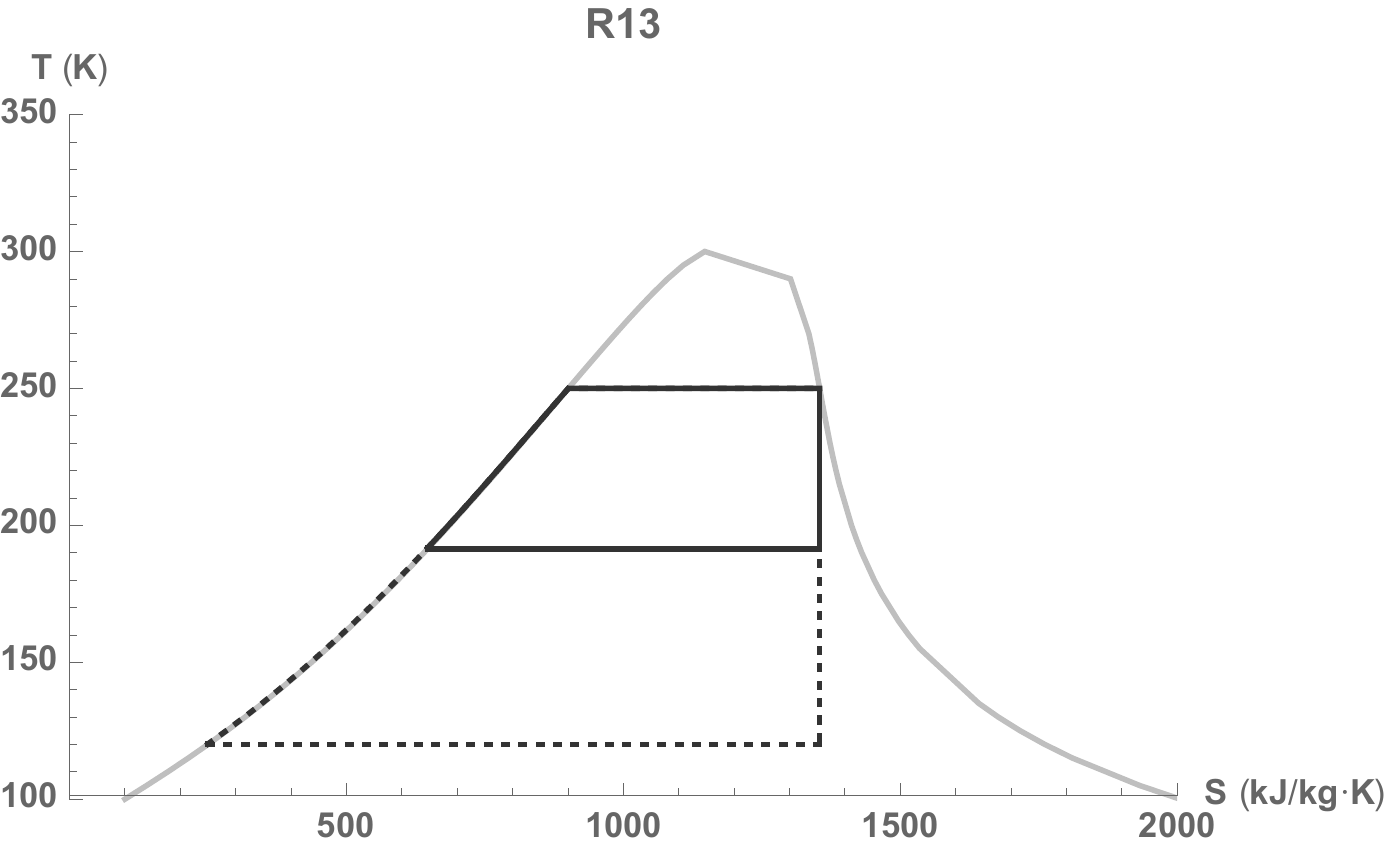}
	\includegraphics[width=\columnwidth]{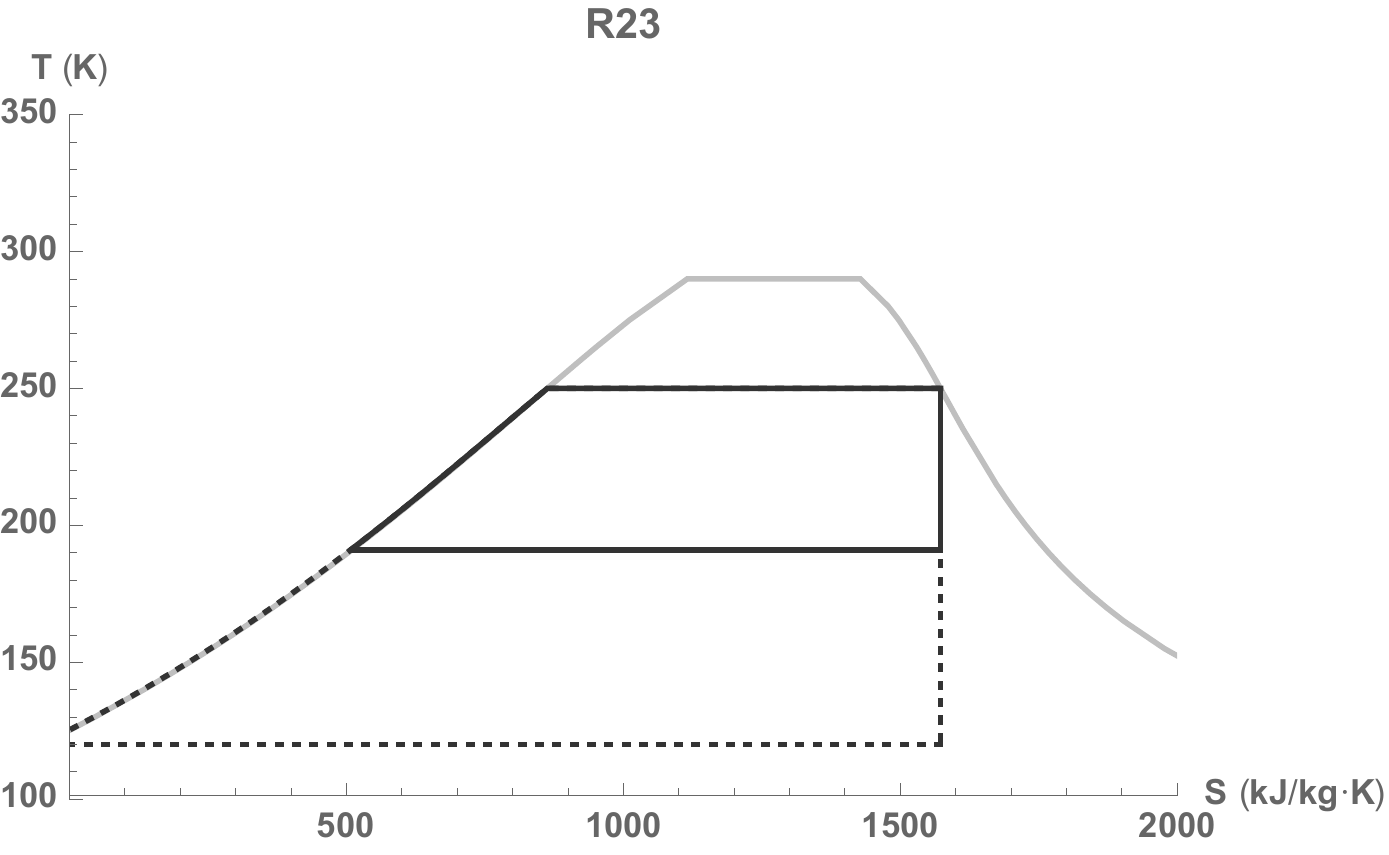}
	\includegraphics[width=\columnwidth]{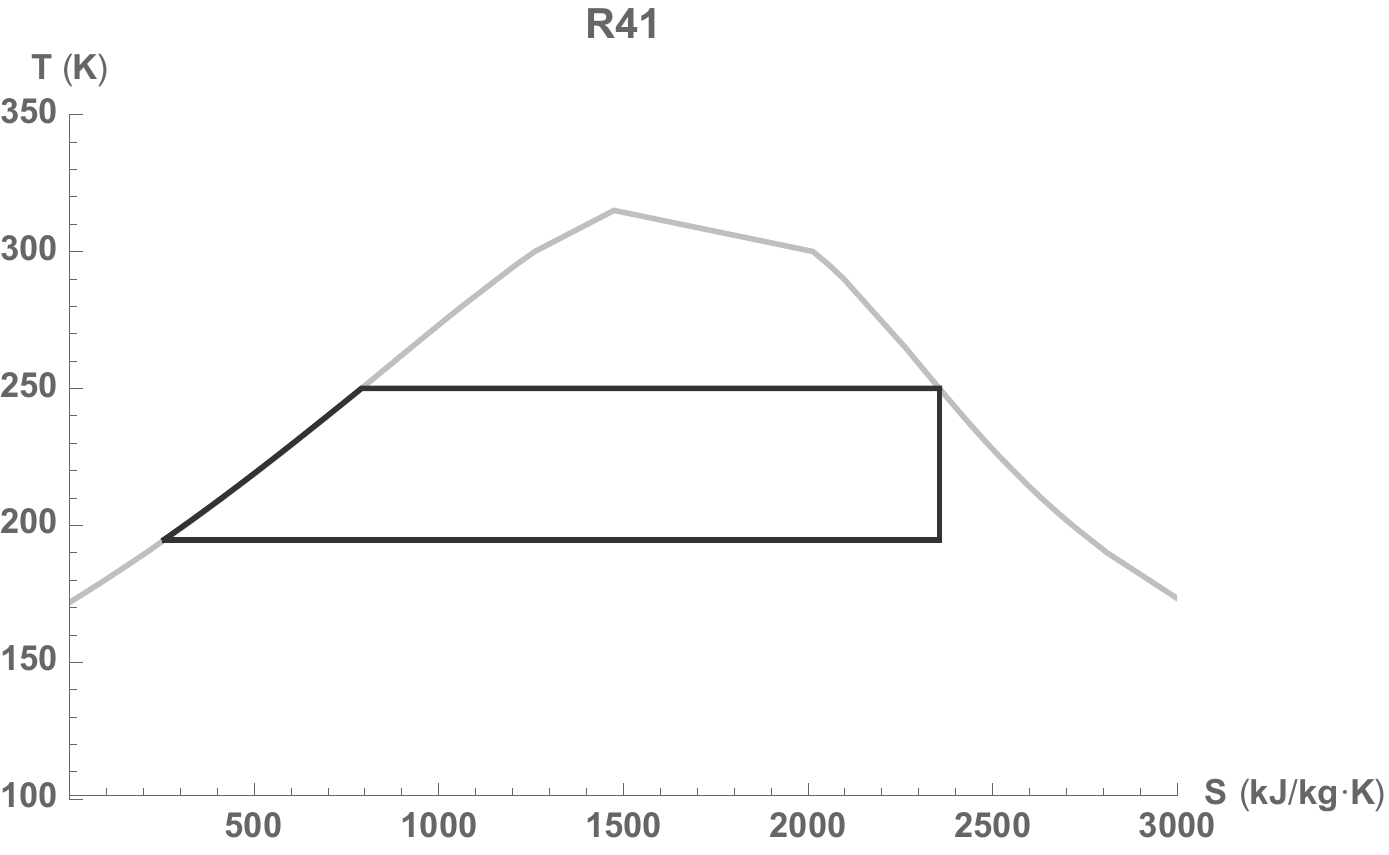}
	\includegraphics[width=\columnwidth]{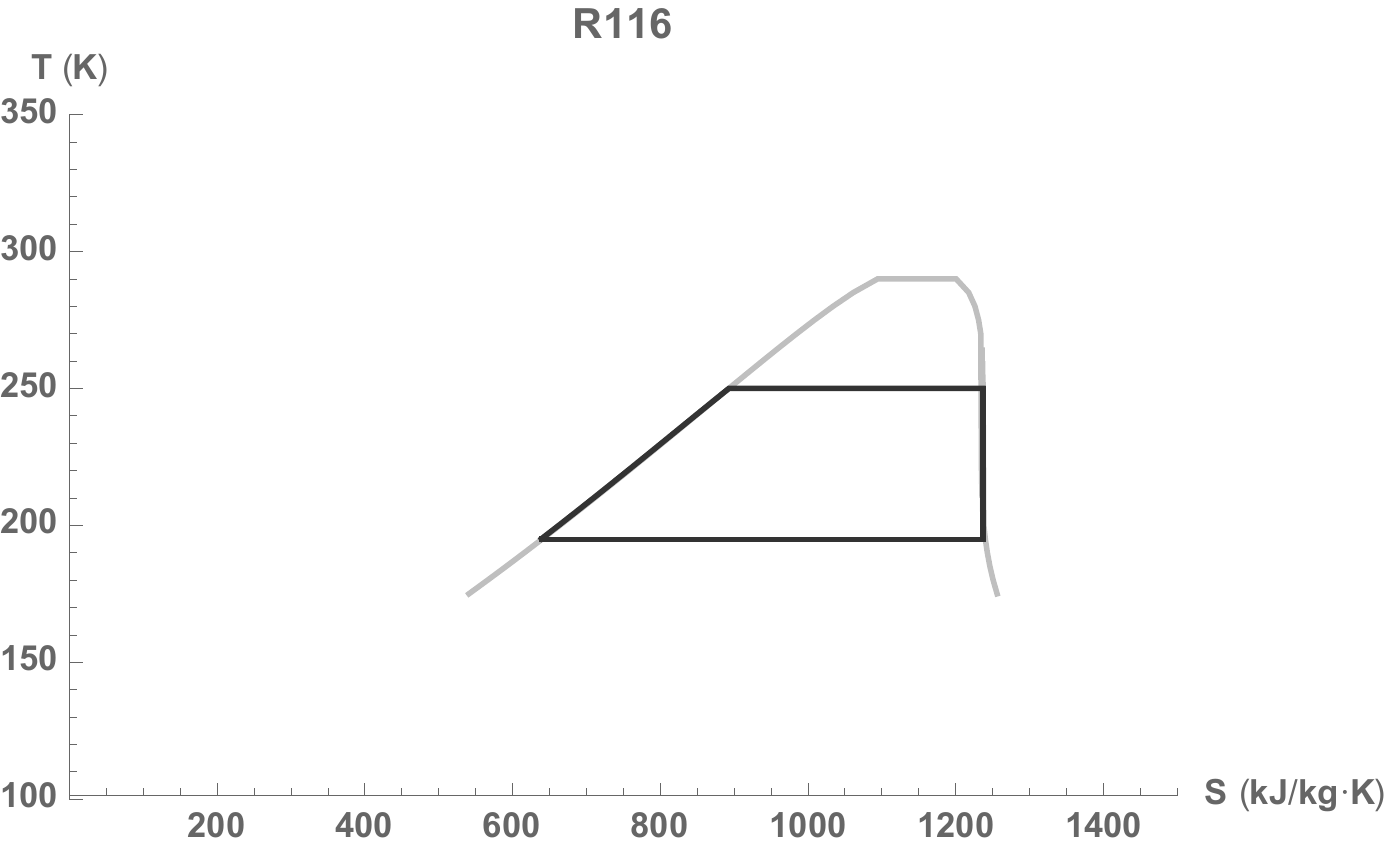}
	\includegraphics[width=\columnwidth]{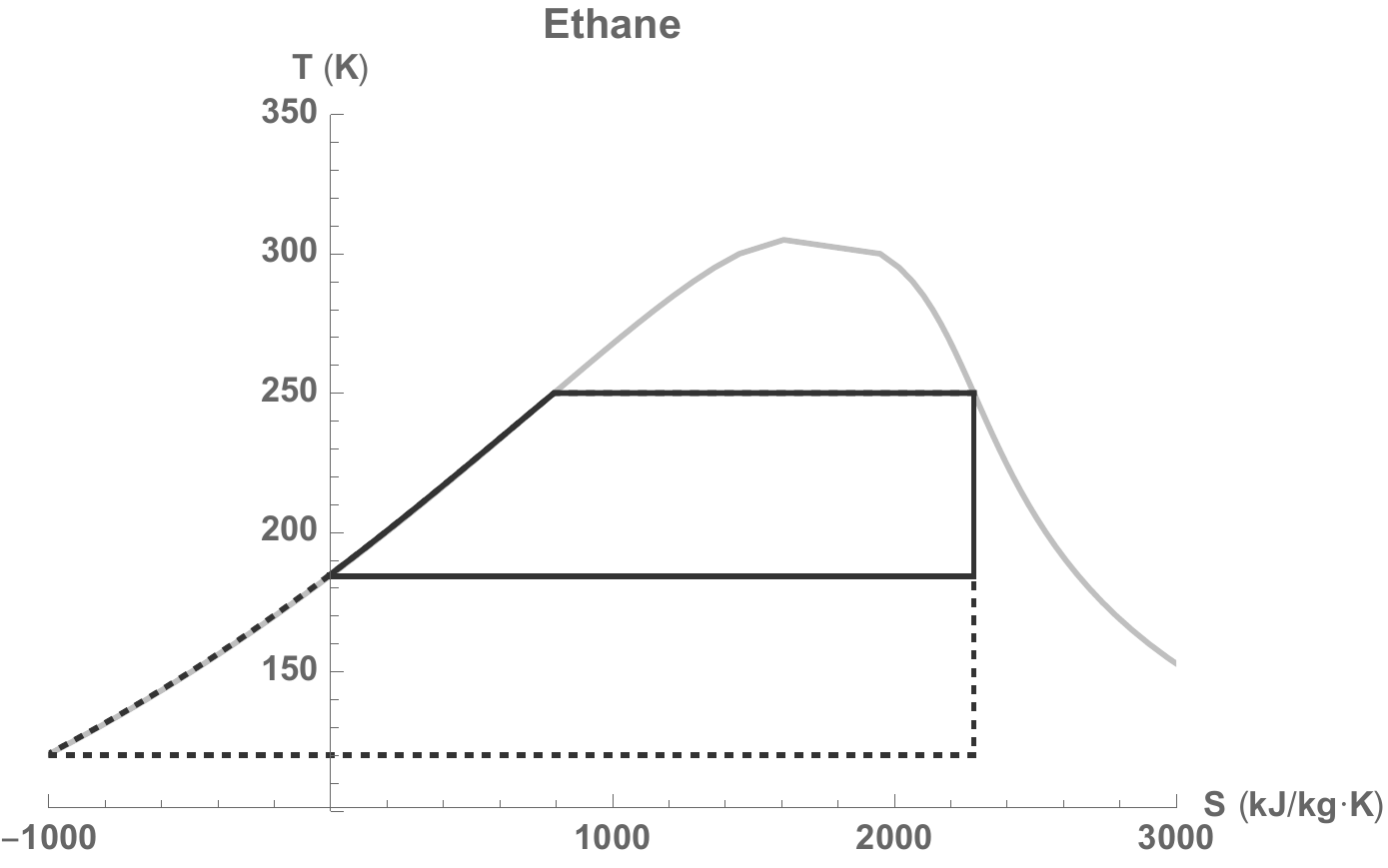}
	\caption{Temperature-Entropy (T-S) diagrams for ORC with candidate working fluids (note different S axis scales). The dotted lines represent the ideal cycles and the full lines the cycles after boiler and condenser pressure constraints are introduced.}
	\label{fig:TSDiagrams}
\end{figure*}

In order to raise the condenser pressure, the heat exchange temperature needs to be raised. In Table\,\ref{tbl:working_fluids_pressure}, we present the resulting cycle parameters and performance figures after introducing the minimum and maximum pressure criteria. The T-S diagrams are represented in Figure\,\ref{fig:TSDiagrams}.

\begin{table*}
	\centering
	\caption{Results for candidate working fluid candidates in the ideal ORC with $T_{\rm H} \leq 250\,{\rm K}$ and $T_{\rm C} \geq 120\,{\rm K}$, with the imposed pressure limits $p_{\rm boil} \leq 2\,{\rm MPa}$ and
	$p_{\rm cond} \geq 0.1\,{\rm MPa}$. The listed temperatures $T_{\rm H}$ and $T_{\rm C}$ listed are the ones after applying pressure limits. Also listed are some relevant parameters for turbine performance, namely, specific volume flow ratio $v_2/v_1$ and outlet vapour quality, $x_2$.}
	\label{tbl:working_fluids_pressure}
	\begin{tabular}{l | c c | c c | c c | c c}

		Fluid	& $p_{\rm boil}$& $p_{\rm cond}$	& $T_{\rm H}$	& $T_{\rm C}$	& $v_2/v_1$	& $x_2$		& $\eta$	& $\dot{W}_{\rm t}/\dot{m}$	\\
		~		& $\rm (kPa)$	& $\rm (kPa)$		& $\rm (K)$		& $\rm (K)$		&			&			&			& $\rm (kJ/kg)$ \\
		\hline
		\hline
		R13		& $1040$		& $100$				& $250$			& $191$			& $8.65$	& $90.6\%$	& $20\%$	& $34.6$ \\
		R23		& $1263$		& $100$				& $250$			& $191$			& $10.0$	& $84.7\%$	& $20.5\%$	& $53.1$ \\
		R41		& $1030$		& $100$				& $250$			& $194.9$		& $7.88$	& $83.7\%$	& $19.8\%$	& $102$ \\
		R116	& $950$			& $100$				& $250$			& $194.8$		& $9.02$	& $99.4\%$	& $18.2\%$	& $26.5$ \\
		Ethane	& $1300$		& $100$				& $250$			& $184.3$		& $10.0$	& $86\%$	& $22.6\%$	& $126$ \\
	\end{tabular}\\
\end{table*}

%-----------------------------------------------------------------------------%

\subsection{Turbine and Pump Efficiencies}

The turbine and the pump are the main sources of entropy increase in the cycle. The turbine, specifically, is a critical and complex design component that demands a wide range of requirements that we shall not explore here, since our aim is to show the feasibility of the ORC in a new field of applications.

One key aspect, even before considering performance, is the reliability and durability of the turbine. Since we have designed our cycle as a saturated vapour cycle, the fluid at the turbine outlet will, in most cases, be a vapour-liquid mixture. The presence of liquid may cause premature erosion of the turbine blades and reduce its efficiency, Typically a vapour quality, $x_2$, of at least $85\%$ at the turbine outlet is desirable \cite{Bao:2013}. As shown in Table\,\ref{tbl:working_fluids_pressure}, all candidate fluids are, at least, close to meeting this criterion. If required, the turbine outlet vapour dryness can be increased by super-heating the fluid beyond saturation. This can be achieved, either by increasing the temperature or reducing boiler pressure.

Another important parameter is the specific volume ratio between the turbine outlet and inlet, $v_2/v_1$. lower volume ratio allows for the use of simpler and cheaper turbines \cite{Saleh:2007,Bao:2013}. The volume flow ratio is also important to achieve a high turbine efficiency. In order to achieve a turbine efficiency over 80\%, the volume flow ratio should be lower than 50 \cite{Macchi:1981,Angelino:1991,Bao:2013}. As also seen in Table\,\ref{tbl:working_fluids_pressure}, we are well below that limit, and thus it is safe to assume that a turbine efficiency around $85\%$ can be obtained. If we consider a similar efficiency in the pump, which would be a conservative estimate \cite{Aljundi:2011}, the cycle performance would be as shown in Table\,\ref{tbl:working_fluids_turbine}, where the required mass flow rate for an initial colony consuming $25\,{\rm kW}$ of sustained power is also computed.

\begin{table*}
	\centering
	\caption{Results for different working fluid candidates in the ORC.}
	\label{tbl:working_fluids_turbine}
	\begin{tabular}{l | c c c | c }

		Fluid	& $\eta$	& $\dot{W}_{\rm t}/\dot{m}$		& $\dot{m}(25\,{\rm kW})$ 	& $A_{\rm rad}$\\
		~		&			& $\rm (kJ/kg)$ 		& $\rm (kg/s)$				& $\rm (m^2)$ \\
		\hline
		\hline
		R13		& $16.9\%$	& $29.4$				& $0.85$ 					& $1745$ \\
		R23		& $17.3\%$	& $45.1$				& $0.554$ 					& $1724$ \\
		R41		& $16.8\%$	& $86.9$				& $0.288$ 					& $1665$ \\	
		R116	& $15.5\%$	& $22.5$				& $1.11$ 					& $1819$ \\
		Ethane	& $19.2\%$	& $107$				& $0.234$ 					& $1746$ \\
	\end{tabular}\\
\end{table*}

%-----------------------------------------------------------------------------%

\subsection{Heat Exchangers}

In order to heat and cool the fluid, some kind of heat transfer apparatus must be set up with both the cold and hot sources. On the surface, cooling of the fluid may be achieved by a radiator, while the heating in contact with the regolith at a few meters depth could use a Ground-Coupled Heat Exchanger (GCHE) \cite{Soni:2015}.

If condensation heat is reject by a radiator, one can estimate the required surface area using
\begin{equation}
	\dot{Q}_{\rm out} = \varepsilon_{\rm rad} A_{\rm rad} \sigma (T_{\rm rad}^4 - T_{\rm back}^4),
\end{equation}
where $\varepsilon_{\rm r}$ is the emissivity of the radiator, $A_{\rm r}$ is its surface area, $\sigma$ is Stefan-Boltzman's constant, $T_{\rm rad}$ is the radiator temperature and $T_{\rm back}$ is the background radiation temperature.

During the lunar night, $T_{\rm back}$ will be the temperature of the radiation coming from outer space, which will be very low and, therefore, negligible. Since heat rejection happens in the phase transition zone, the temperature of the fluid should remain constant and we can take $T_{\rm rad} \simeq T_3$ as a figure of merit. In theory, given the very low background temperature, $T_{\rm rad}$ could be made as low as intended by increasing the radiation area for a given heat flow. However, is is unlikely that any radiator can be completely insulated from the surface, so we take surface temperature.

Assuming $T_{\rm back} = 4\,{\rm K}$ and an emissivity $\varepsilon_{\rm rad} = 0.9$, we have computed the radiator are required for a $25\,{\rm kW}$ and present the results in Table\,\ref{tbl:working_fluids_turbine}. Of course, the heat from these radiators might be used for other purposes such as, for instance, heating the habitable areas.

%-----------------------------------------------------------------------------%
%-----------------------------------------------------------------------------%

\section{Conclusion}

We have shown that a power system based on an Organic Rankine Cycle is a viable option to power a permanent lunar installation, either manned or automated. Using concepts already used on Earth, taking advantage of existing temperature gradients in the lunar regoltih, a thermal engine can be powered. We have also considered a set of possible working fluids to power this engine, evaluating their performance in terms of thermal efficiency and net work per mass flow rate.

This opens up new options when considering lunar settlement or long term exploration and exploitation of resources on the Moon. Our proposal is particularly relevant and realistic to ensure that there is no interruption of supply during the 14-day long night that must be endured on the Moon and when conventional solar photovoltaic power is not available.

%-----------------------------------------------------------------------------%
%-----------------------------------------------------------------------------%

\section*{acknowledgements}

The work of FF is supported by the Fundação para a Ciência e Tecnologia through grant SFRH/BPD/118649/2016.

%-----------------------------------------------------------------------------%
%-- Bibliography -------------------------------------------------------------%
%-----------------------------------------------------------------------------%

\bibliographystyle{elsarticle-num}

\bibliography{rankine_moon}

\end{document}